# Research of X-ray induced conductivity of ZnSe sensors for their application in isotopic thickness gauges

A.O. Sofiienko *, V.Ya. Degoda **

* 03680, Ukraine, Kyiv, 152 Gor'kogo St., off. 130, "Ukratominstruments" Corporation, tel./fax: (38) 0445016178, asofienko@gmail.com

** 01601, Ukraine, Kyiv, 64 Volodymyrs'ka St., Taras Shevchenko National University of Kyiv,
tel.: (38) 066 2431530, degoda@univ.kiev.ua

**Abstract**
Measurements of intrinsic conductivity and X-ray induced conductivity were performed on specially undopped ZnSe samples. The measurements demonstrated that sensors made of ZnSe have minor intrinsic conductivity when heating up to the temperature of 180 °C, and significant X-ray induced conductivity. Dose dependence "dose rate - current" is described with simple power function which considerably simplifies calibration of sensors. This results can be used during the designing of high-temperature X-ray and gamma-radiation detectors for radiation hot rolling thickness gauges which are widely used in the metallurgy.

## 1. Introduction

Radio isotopic devices are being successfully used in many process control systems for several decades already. No alternative has been found for many methods based on the use of ionizing radiation. Bright proof of this statement is the material radio isotopic thickness gauges (ITG). ITG are used in many branches of industry [1-3]. Although, most of the sensors are used in the metallurgy, for example, in automatic metal thickness control of rolling mill. Among the variety of radiation detection methods known nowadays scintillation and gas-discharge methods are mainly used in radiation thickness gauges. Each of these methods of ionizing radiation registration has its own drawbacks. For example there are limitations on heating temperature for scintillation materials, and gas-discharge sensor based systems do not allow obtaining high spatial and time resolution. Semiconductor wide-band materials having high temperature and radiation resistance can be used as the alternative in the detectors of radiation thickness gauges. Specially undopped single-crystal ZnSe is suggested to be used as a future-technology material for high temperature ionizing radiation detectors, this material has high radiation resistance which has been proven in series of researches [4, 5]. An objective of present research is to study electrical conduction of ZnSe sensors at high temperatures and high power X-ray fields.

## 2. Methods of researches

Conductivity of ZnSe samples was analyzed by means of excitation using X-ray quanta. Specially undopped ZnSe crystals had been grown after preliminary cleaning of the burden in order to obtain crystals with minimum concentration of admixtures and maximum specific resistance ($10^{11} - 10^{12}$ Ohm·cm). To study conductivity multilayer metal contacts were sputtered on the crystals using resistive method; and conductors were soldered to these contacts. Quality of contacts was verified by measuring volt-ampere characteristics of X-ray induced conductivity and intrinsic conductivity of the samples. The distance between electrodes was 5 mm. The voltage up to 1500V was supplied to one electrode, and another electrode was grounded through nanoamperemeter. Experimental researches were conducted within the temperature range from 85 K up to 513 K (-190 ÷ +240 $^0$C). X-ray excitation was done by means of total radiation of X-ray tube BHV7 (Re - anode, $U_T$ = 20 kV, $i_T$ = 5÷25 mA) through cryostat beryllium window. Calculated exposure dose rate varied within the range of 1.8·10⁵ R/h up to 9·10⁵ R/h in the plane of sample. During the research residual pressure maintained at the level of $10^{-3}$ mm of mercury column inside the cryostat in order to decrease probability of electrical breakdown on the surface of the samples when voltage on the electrodes is increased up to 1500 V. Since X-ray radiation was used as an exciting radiation which is absorbed completely in the thickness of ZnSe 80 mkm ($E_X$ = 20 keV), then geometry of electrodes location on the sample surface was selected (Fig. 1).

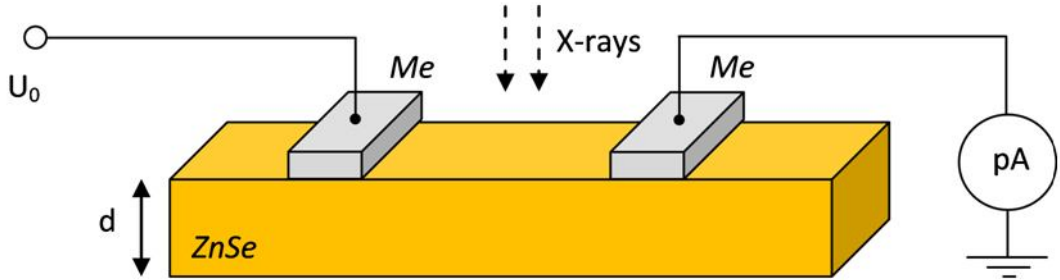

Fig. 1 Geometry of current measurement of X-ray induced conductivity of ZnSe samples
($U_0$ = 0-1500 V; d = 2-5 mm)

The area between electrodes was uniformly irradiated with X-ray quanta flux from X-ray tube. Such geometry corresponds to the case when high-energy gamma radiation (660 – 1300 keV) is uniformly absorbed along the depth in flat detector having active area 300-500 mkm.

Exposure dose rate of X-ray radiation in the sample plane was evaluated using calculated method [6]:

$$D_X \approx 5.1 \cdot 10^{-2} \cdot \Phi_X \cdot \overline{E_X} \cdot \mu_a(\overline{E_X}) \cdot \exp(-d_{Al} \cdot \mu_{Al}(\overline{E_X})), \text{ R/h} \qquad (1)$$

where $\Phi_X$ – flux of X-ray quanta in the plane of source (1/s·cm²); $\overline{E_X}$ – mean energy in tube radiation spectrum (MeV); $\mu_a$ – linear coefficient of X-ray radiation true absorption in the air (cm⁻¹); $d_{Al}$ – thickness of outlet window of X-ray tube brought to the equivalent thickness of aluminum screen (0.04 cm); $\mu_{Al}(\overline{E_X})$ – linear coefficient of X-ray radiation attenuation having energy of $\overline{E_X}$ = 10 keV in the outlet window of X-ray tube (71.28 cm⁻¹). Quanta flux in the sample plane is determined with the radiation power of X-ray tube:

$$\Phi_X = \frac{2P_T}{2\pi \cdot r^2 \cdot e_k U_T} = \frac{k_0 \cdot Z_T \cdot i_T \cdot U_T}{\pi e_k \cdot r^2}, \qquad (2)$$

where $k_0$ = 10⁻⁹ (B⁻¹); $Z_T$ – ordinal number of the substance of X-ray tube anode (Re); $i_T$ – filament current of X-ray tube; $U_T$ – accelerating potential of X-ray tube; $e_k$ = 1.6·10⁻¹⁹ coulomb; $r$ – distance from X-ray tube anode to sample plane (15 cm). Considering the geometry where ZnSe samples were irradiated the following relation was used to evaluate the value of exposure dose rate in the plane of samples: $D_X \approx 3.6 \cdot 10^7 \cdot i_T$ (R/h).

## 3. Results and discussion

Research of temperature characteristics of intrinsic conductivity of single and polycrystalline ZnSe samples shows that they have dominating centers of thermoactivated conductivity of n-type, and in coordinates $\ln(I_D) - T^{-1}$ conductivity curve is described precisely enough with linear function. Conductivity type of specially undoped ZnSe samples is defined with thermo-voltage method. Thermo activated centers have energy position $\Delta E \approx 0.82$ eV for polycrystal samples. When heating polycrystalline samples up to the temperature of $T \geq 100\ ^0C$ the value of intrinsic conductivity is compared by the value to X-ray induced conductivity given the specified operation modes of X-ray tube.

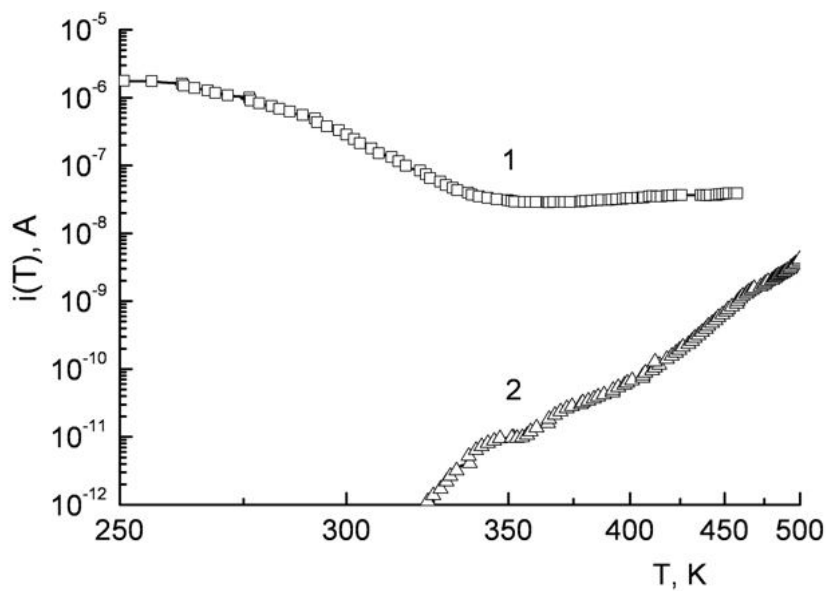

Fig. 2 Temperature dependencies of X-ray induced conductivity of single-crystal ZnSe sample (1) and its intrinsic conductivity (2), $E_0 = 400$ V/cm

Single-crystal samples are characterized with the thermoactivated centers having energy position $\Delta E \approx 1.1$ eV. Temperature dependence of X-ray induced conductivity of single-crystal ZnSe and intrinsic conductivity is shown in Fig. 2. Calculated exposure dose rate in sample plane was as follows during measurements of X-ray induced conductivity $D_X \approx 0.72$ MR/h.

Obtained results show that under the control of temperature of single-crystal sample of ZnSe total current of the sensor can be corrected to the intrinsic conductivity current using known calibration functions of intrinsic conductivity of the sample at different temperatures. Polycrystalline samples cannot be used in the sensors of ionizing radiation at high temperatures due to extremely high intrinsic conductivity which is stipulated by high concentration of intrinsic defects of crystal lattice, and as a consequence low active energy of thermoativated conductivity levels.

Important characteristic of the sensor of ionizing radiation is linear dependence of its charge or current response on the value of the flux of registered radiation. Current of X-ray induced conductivity was measured at different values of dose rate of X–ray tube radiation and different bias voltages on sample electrodes ($U_0$) in order to define linearity of current response of sensors from single-crystal ZnSe. The results are shown in Fig. 3. It was defined that when bias voltage is increased up to 600 V ($E_0 = 1200$ V/cm) the dependence of voltage of X-ray induced conductivity on the dose rate is over-linear and described with the following function $i_X \sim D^\beta$. When the densities of the field are significant ($E_0 > 1400$ V/cm) the dependence of current of X-ray induced conductivity on the dose rate is close to the linear (relevant dependence is given in Fig.4). When the duration of measurements is less than 6 hours and value of absorbed dose is more than 0.3 Mrad stability of voltage-current characteristic of the sample is pointed-out.

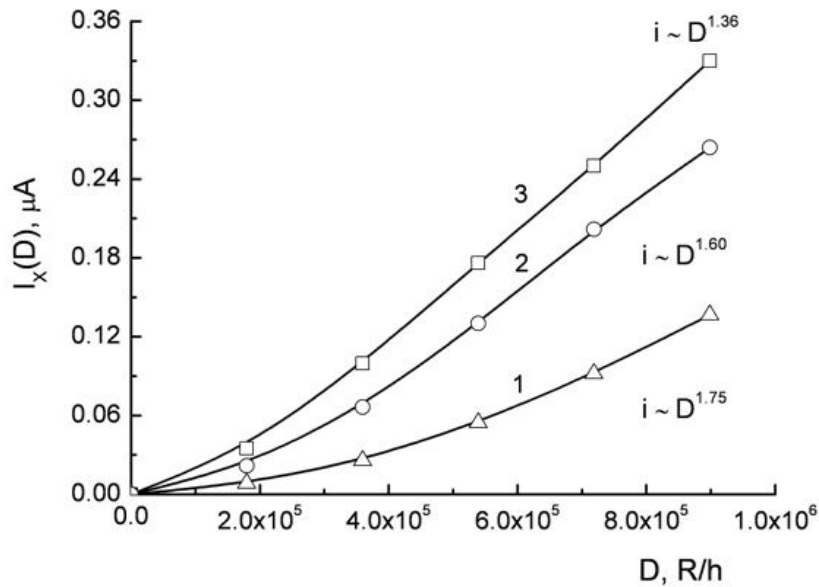

Fig. 3 Dependence of current of X-ray induced conductivity of ZnSe sensor on the dose rate of X-ray radiation at T = 295K and different bias voltages:
200 V (1); 400 V (2); 600 V (3)

Thus, it was defined that when the values of density of electric field are $E_0 > 1400$ V/cm stability of sensors operation is increased in the samples of single-crystal ZnSe, the "dose rate – current" dependence is linearized, and calibration procedure on the dose rate becomes simpler.

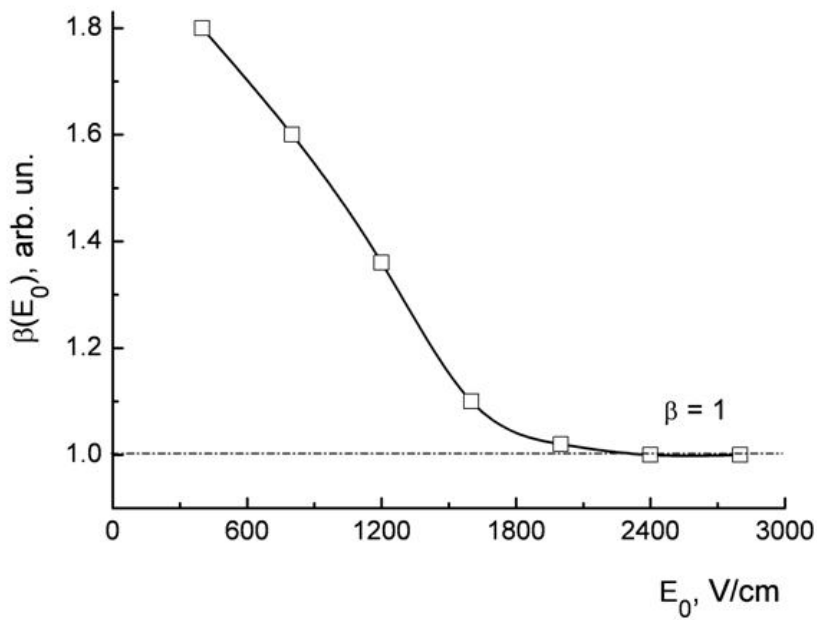

Fig. 4 Dependence of exponent of dose characteristic function ($i_X \sim D^\beta$) on the density of electric field in ZnSe samples

It is necessary to note that depending on the batch of ZnSe samples the values of their specific resistance can vary due to the change of concentration of non-controlled impurity centers, crystal defects which leads to the increase or decrease of charge response by several times at constant level of photon radiation. To select optimal sensor it is necessary to perform special control of electric characteristics of ZnSe samples within wide range of temperatures from +10 $^0$C up to +240 $^0$C.

## 3. Conclusion

Change of intrinsic conductivity of ZnSe samples within the range of temperatures from 10 $^0$C up to 240 $^0$C showed that single-crystal specially undoped ZnSe has extremely low intrinsic conductivity. This attribute of obtained samples can be used during the designing and manufacturing of gamma and X-ray radiation detectors for the application in radiation hot rolling thickness gauges which are widely used in the metallurgy. Distinctive feature of such detectors is that there is no necessity to perform additional cooldown which considerably simplifies measuring part of thickness gauge. It was determined that "dose rate - current" calibrating characteristic of analyzed samples approaches to the linear one when electric field is increased in the sample up to $E_0 \geq 1400$ V/cm. However, since the current of X-ray induced conductivity is described precisely enough with simple power function of the following type $i_X \sim D^\beta$, which is linearized in double logarithmic scale, then in practice significantly lesser electric fields 500-1000V/cm can be used what decreases the probability of surface breakdown of sensors.

## References

[1] Rothwell R., 1973. Nucleonic thickness gauges - a SAPPHO pair. Research Policy, 2, No. 2, 144-156.
[2] Artemiev B.V., 2002. X-ray Metal Thickness Measuring. Mashinostroenie-1, 102-105.
[3] Sowerby B.D., Rogers C.A., 2005. Gamma-ray density and thickness gauges using ultra-low activity radioisotope sourcesfillin. Applied Radiation and Isotopes, 5, No. 5-6, 789-793.
[4] Ryzhikov V., Starzhinskiy N., Gal'chinetskii L. et al., 2001. Behavior of new ZnSe(Te,O) semiconductor scintillators under high doses of ionizing radiation. IEEE Trans. Nucl. Sci., 48, No. 4, 1561-1564.
[5] Focshaa A. A., Gashina P. A., Ryzhikovb V. D., Starzhinskiy N. G., Gal'chinetskiib L. P. and Silin V. I., 2002. Properties of semiconductor scintillators and combined detectors of ionizing radiation based on ZnSe(Te,O)/pZnTe–nCdSe structures. Optical Materials, 19, No. 1, 213-217.
[6] Luyanov V.B., Berdonosov S.S., 1977. Radioactive indicators in Chemistry. Higher School, Moscow.